\documentclass{emulateapj}
\usepackage{apjfonts}
\usepackage{epsf}
\usepackage{amsmath}

\shorttitle{And XXIX: A New Andromeda Satellite}
\shortauthors{Bell, Slater \& Martin}

\slugcomment{{\sc The Astrophysical Journal Letters}: in press }

\begin{document}


\title{Andromeda XXIX: A New Dwarf Spheroidal Galaxy 200\,kpc from Andromeda}

\author{Eric F.\ Bell and Colin T.\ Slater }
\affil{Department of Astronomy, University of Michigan,
    500 Church St., Ann Arbor, MI 48109}

\and

\author{Nicolas F.\ Martin}
\affil{Max-Planck-Institut f\"{u}r Astronomie, K\"{o}nigstuhl 17,
    D-69117 Heidelberg, Germany}

\begin{abstract}
We report the discovery of a new dwarf galaxy, Andromeda XXIX, using data from
the recently-released Sloan Digital Sky Survey DR8, and confirmed by Gemini North telescope Multi-Object Spectrograph imaging data.  And XXIX appears to be a dwarf spheroidal galaxy, separated on the sky by a little more than 15 degrees from M31, with a distance inferred from the tip of the red giant branch of $730\pm75$\,kpc, corresponding to a three dimensional separation from M31 of $207^{+20}_{-2}$\,kpc (close to M31's virial radius).  Its absolute magnitude, as determined by comparison to the red giant branch luminosity function of the Draco dwarf spheroidal, is $M_V = -8.3\pm 0.4$.  And XXIX's stellar populations appear very similar to Draco's; consequently, we estimate a metallicity for And XXIX of [Fe/H]$\sim -1.8$.  The half-light radius of And XXIX is $360\pm60$\,pc and its ellipticity is $0.35\pm0.06$, typical of dwarf satellites of the Milky Way and M31 at this absolute magnitude range.
\end{abstract}


\section{Introduction}

The process of building a more complete understanding of the dwarf galaxy luminosity function in the Local Group has been an important activity over the last decade.  Motivated in great part by the seeming discrepancy between the predicted slope of the dark matter subhalo mass function and the apparently flatter slope of the dwarf galaxy luminosity (or velocity dispersion) function (\citealp{klypin99}, \citealp{moore99}; see e.g., \citealp{maccio10} or \citealp{font11} for discussion of relevant considerations for resolving this apparent discrepancy), dwarf galaxy searches using modern survey datasets have multiplied significantly the number of known Local Group galaxies.  Around the Milky Way, the Sloan Digital Sky Survey (SDSS; \citealp{SDSS}) has been the key dataset \citep[e.g.,][]{willman1,belokurov07}.  Around M31, the bulk of the new dwarf discoveries have come from dedicated surveys with the Isaac Newton Telescope \citep{irwin08} and the Canada-France-Hawaii Telescope \citep{ibata07, mcconnachie09, martin06, martin09}. These surveys have obtained deep observations over a significant fraction of the area within $\sim$150 kpc of Andromeda.  In addition to these dedicated surveys, two satellites of Andromeda
have been found in an early SDSS imaging scan targeting Andromeda \citep[And IX and X,][]{zucker04,zucker07}.  

The goal of this Letter is to report the discovery of a new dwarf spheroidal galaxy, Andromeda XXIX\footnote{Though the dwarf is located within the constellation
Pegasus, probable satellites of Andromeda are by convention named and numbered
with the prefix Andromeda. See \citet{martin09}.} in area recently released by the SDSS in their Data Release 8 \citep[DR8,][]{DR8}.  The SDSS DR8 includes imaging coverage of $\sim 3200$ deg$^2$ in the south Galactic cap, allowing access to nearly half of the area within $35^\circ$ ($\sim 450$ projected kpc) of Andromeda, extending the radial range over which dwarf galaxies can be discovered.  While the SDSS is substantially shallower than the
dedicated M31 surveys, it is deep enough to enable discovery of dwarf galaxies down to luminosities of $M_V \sim -8$ (see \citealp{slater11} for discussion of And XXVIII, a slightly brighter and more compact galaxy discovered in the SDSS DR8).

\section{Detection}

\begin{figure}
\begin{center}
\plotone{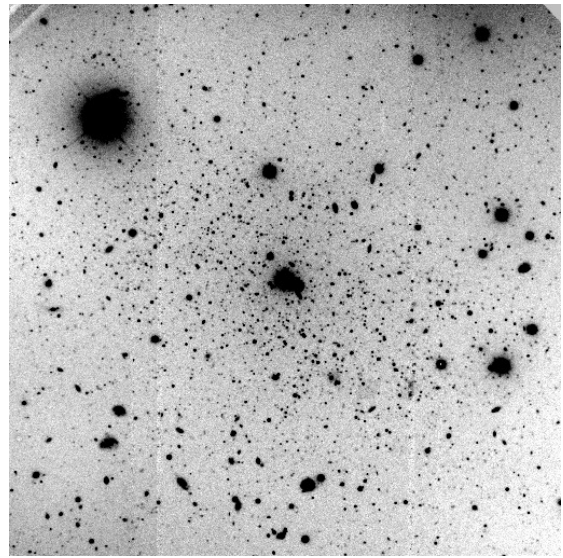}
\caption{A 500 second $r$-band image from GMOS-N of And XXIX.  The image is
$5^\prime \times 5^\prime$.  North is up, and east is to the left.  The bright sources in the middle are an $r=14.13$ star (with SDSS colors placing the detection right on the $ugr$ stellar locus) blended (at this stretch) with two fainter $19<r<21$ sources.
\label{fig:image}} 
\end{center}
\end{figure}

\begin{figure*}
\begin{center}
\plotone{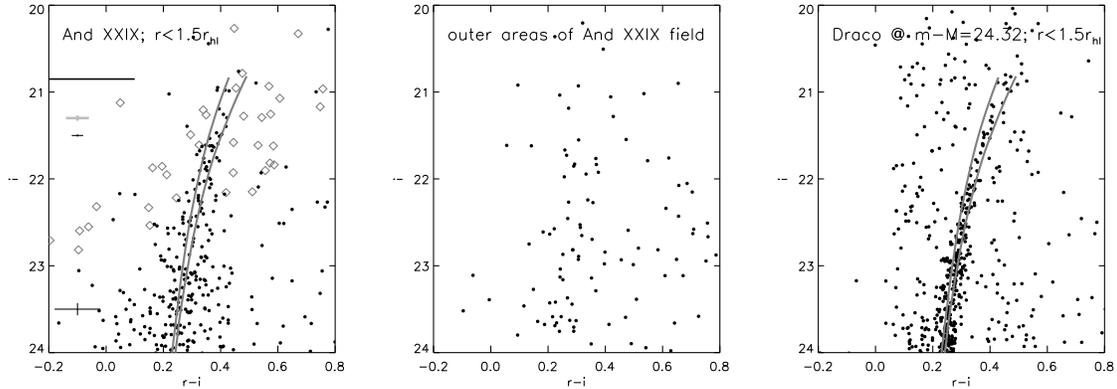}
\caption{{\it Left:} the color--magnitude diagram of DAOPHOT-measured stars from GMOS-N imaging (solid points) of the region within 1.5 half-light radii around And XXIX.  Typical random photometric uncertainties for the DAOPHOT-measured stars are shown as black error bars, and the calibration uncertainty is shown as the light grey error bar.  Diamonds show the SDSS colors and magnitudes of stars within 2$^{\prime}$ of And XXIX.  The thick black horizontal line shows the approximate TRGB magnitude, implying a distance modulus of 24.32$\pm0.22$ ($730\pm 75$\,kpc).  The two gray lines show the \citet{dotter08} isochrone for [Fe/H]$= -2$ (left) and $-1.5$ (right), both with scaled solar abundance ratios and 12\,Gyr age.   {\it Center:} the color--magnitude diagram of stars more distant from And XXIX drawn from a similar area.  A weak but clear RGB is visible, as the field of view of GMOS-N is too small to give a completely clean control sample.  {\it Right:} the color--magnitude diagram of the central 1.5 half-light radii of Draco (from \citealp{segall06}), scaled in brightness to match the distance modulus of And XXIX.  
\label{fig:cmd}}
\end{center}
\end{figure*}

\begin{table}
\caption{Properties of And XXIX \label{properties_table}}
\begin{center}
\begin{tabular}{l c}
\tableline
Parameter & \\
\tableline 
\tableline
$\alpha$ (J2000) & 23\fh 58\fm 55\fs 6  \\
$\delta$ (J2000) & 30\arcdeg 45\arcmin 20\arcsec \\
$E(B-V)$ & 0.048\tablenotemark{1} \\
$(m - M)$ & $24.32 \pm 0.22$ \\
$D$ & $730\pm 75$\,kpc \\
$r_{\rm M31}$ & 207$^{+20}_{-2}$\,kpc \\
$M_V$ & $ -8.3 \pm 0.4$ \\
Ellipticity & $0.35 \pm 0.06$ \\
Position angle (N to E) & $51 \pm 8$ degrees \\
$r_h$ & $1.7^{\prime} \pm 0.2^{\prime}$ \\
$r_h$ & $360 \pm 60$\,pc \\
${\rm [Fe/H]}$ & $\sim -1.8$\tablenotemark{2} \\
\tableline
\tablenotetext{1}{\citet{SFD98}} 
\tablenotetext{2}{Given the great similarity between the CMDs of Draco and And XXIX, we estimate the metallicity to be the same as that of Draco, adopting the metallicity estimate of \citet{apparicio01}.}
\end{tabular}
\end{center}
\end{table}

At the distance of Andromeda (785\,kpc; \citealp{mcconnachie05}; this is consistent with the mean RGB, Cepheid and RR Lyrae distances tabulated in the NASA/IPAC Extragalactic Database, but is somewhat more distant than recent eclipsing binary determinations; e.g., \citealp{vil10}), 
searches for dwarf galaxies in SDSS are limited to using red giant branch (RGB)
stars as tracers of the underlying population of main-sequence and subgiant
stars. Alternative tracers commonly used for detecting dwarf galaxies around the
Milky Way, such as the horizontal branch or the main sequence turn-off, are much
too faint to be detected.  

To search for dwarf galaxies in the SDSS we compute star
counts in $2^\prime \times 2^\prime$ bins, selecting only stars with $0.3 < 
r-i < 0.8$, colors roughly similar to metal-poor giant branch stars.
Overdensities are readily apparent upon visual inspection of the resulting map
as ``hot pixels'', typically with counts of 10--15 objects (stars and unresolved galaxies) as compared to the
background of 1--3 objects per bin.  With this technique, we were able to detect many known dwarf companions of M31 (e.g., And I, II, III, IX, XXII, LGS3, Pegasus Dwarf Irregular galaxy; see also \citealp{richardson11}).  However, 
most of the detected overdensities 
were galaxy clusters
at intermediate redshift, which have similar colors and contain many
spatially-unresolved member galaxies. Visual inspection of the SDSS image along
with the color-magnitude diagram (CMD) is sufficient to reject the majority of the 
false-positives.

Andromeda XXIX appeared as a marginally significant detection, and deep follow-up data were obtained through Director's discretionary time (proposal ID GN-2011A-DD-6) at Gemini North, with the Gemini North telescope Multi-Object Spectrograph (GMOS-N; \citealp{gmosn}).  Images in $r$ and $i$ bands were taken at two dither positions, for a total exposure time of 500 seconds in each band.  The data were reduced using the standard Gemini pipeline in IRAF, including de-bias and flat fielding using twilight flats.  The FWHM is $\sim 0.6 \arcsec$ in each band, and the data are calibrated to $\sim 0.03$\,mag accuracy through cross-calibration with SDSS stars between 18th and 20th magnitude (in the band of interest) in the Gemini field.  The $r$-band image of And XXIX is shown in Fig.\ \ref{fig:image}, and the properties of And XXIX are summarized in Table~\ref{properties_table}. 

\section{Properties of And XXIX}

The CMD of stars (as measured by DAOPHOT; \citealp{stetson87}; crosses) from the GMOS-N imaging of the region within 1.5 half light radii of And XXIX is shown in the left-hand panel of Fig.\ \ref{fig:cmd}, and that of stars more distant from And XXIX drawn from a similar area in the center panel (corrected for galactic foreground extinction following \citealp{SFD98}).  The left-hand panel clearly shows the upper part of the RGB of And XXIX, and the center panel shows a much weaker but clear RGB (owing to the relatively small field size of GMOS --- the dwarf extends to the edge of the image, but with much reduced stellar density) superimposed on a reasonable approximation to the expected fore- and background contamination in the left panel.  In the left-hand panel, gray diamonds show SDSS colors and magnitudes of stars within 2$^{\prime}$ of And XXIX.  The SDSS data show a very wide color spread, with objects spread across the color range $-0.2<r-i<0.6$, and a higher concentration at $0.1<r-i<0.6$, unlike the narrow RGB seen in the Gemini data.  This is the result of photometric error of the SDSS at such faint levels (at $i \sim 21.25$, we have estimated the photometric error of the SDSS using repeat imaging of SDSS stripe 82, demonstrating uncertainties $\delta i \sim 0.13$\,mag and $\delta(r-i) \sim 0.20$\,mag; \citealp{bramich08}).    

Given that the tip of the red giant branch (TRGB) is the only observable feature in the CMD of And XXIX, and that the TRGB has a roughly constant absolute magnitude in metal-poor stellar systems \citep{bellazini01} of $M_I = -4.04$, we have used it to estimate the distance to And XXIX.  Such a method is commonly applied to dwarf galaxies \citep[e.g.,][]{mcconnachie05,martin09}, but suffers from the disadvantage that the luminous part of the RGB can be very poorly populated owing to the small number of stars in many of the new dwarf discoveries (see, e.g., \citealp{willman1} for a particularly clear example of this problem).  

We used
the maximum-likelihood estimator described in \citet{makarov06}, which modeled
the TRGB luminosity function as:
\begin{equation}
\psi = \begin{cases}
10^{a(m - m_{\rm TRGB}) + b} & m - m_{\rm TRGB} \ge 0, \\
10^{c(m - m_{\rm TRGB})} & m - m_{\rm TRGB} < 0.
\end{cases}
\end{equation}
This broken power-law form takes three parameters: $a$ and $c$ are the slopes of
the luminosity function fainter and brighter than the TRGB, while $b$ is the
strength of the transition at the TRGB. We adopted the values from
\citet{makarov06} of $a = 0.3$ and $c=0.2$, and $b=0.6$. For the TRGB fitting,
the SDSS photometry was converted to Johnson $I$-band using the prescriptions of
R.\ Lupton\footnote{{\scriptsize http://www.sdss.org/dr5/algorithms/sdssUBVRITransform.html\#Lupton2005}}. We find that the likelihood is maximized at
$m_{I,{\rm TRGB}} = 20.28 \pm 0.19$ (shown by the thick black horizontal line in the left-hand panel of Fig.\ \ref{fig:cmd}), and a second, 
less likely (by a factor of four) peak also appears at $m_{I,{\rm TRGB}}=20.83 \pm 0.13$ (in the Gemini $i$-band
$\sim 20.8$ and $\sim 21.3$ respectively).  The reason for this is clear: owing to the small number of RGB stars, there is a clump of stars with $m_i \sim 21$ and a less populated gap, and the number of stars picks up again at $m_i \sim 21.4$.  This gap is almost certainly the result of shot noise in the luminosity function of RGB stars, and we adopt a maximum-likelihood estimate of the distance modulus of $m-M = 24.32 \pm 0.22$ ($730 \pm 75$\,kpc), adding the uncertainty in the TRGB magnitude of 0.12\,mag in quadrature \citep{bellazini01}.  

The two grey lines overlaid on the CMD of And XXIX are the \citet{dotter08} isochrones for [Fe/H]$= -2$ (left) and $-1.5$ (right), both with scaled solar abundance ratios (the TRGB colors depend on alpha enrichment at the $<0.01$ mag level) and 12\,Gyr age, placed at a distance modulus of $m-M = 24.32$.   And XXIX shows a clear and well-defined RGB, clearly different from that of the outer regions of the And XXIX GMOS-N image.  The fact that the isochrones match so well the observed CMD of And XXIX argues that the distance modulus adopted in this letter for And XXIX is not dramatically in error.

As a further check on the properties and estimated distance of And XXIX, the right-hand panel of Fig.\ \ref{fig:cmd} shows the galactic foreground extinction-corrected CMD of stars in the Draco Dwarf Spheroidal from \citet{segall06}, measured using the CFHT Megacam and color-corrected to the SDSS system, shifted to the distance modulus of 24.32, where we have assumed a distance modulus for Draco of 19.59$\pm$0.1, following the RR-Lyrae derived distances of \citet{kinemuchi08} and the compilation of \citet{tammann08}.  One can see that the CMD of And XXIX (including the TRGB region) is very similar to that of Draco (when distance offsets are accounted for).  From this similarity we suggest that And XXIX has a metallicity similar to that of Draco with [Fe/H]$\sim -1.8$ dex (adopting this metallicity from \citealp{apparicio01}).  Furthermore, given the similarity between And XXIX and Draco's CMDs, and that there are no blue luminous stars, such as are in the Leo T dwarf spheroidal galaxy \citep[e.g.,][]{dejong08}, we feel comfortable to classify And XXIX as a dwarf spheroidal galaxy.  

The projected angle between And XXIX and M31 is 15.14 degrees, corresponding to a projected distance of 205\,kpc (assuming M31's distance of 785\,kpc as before).  Given the distance estimate for And XXIX of $\sim 730 \pm 75$\,kpc, the three-dimensional distance of And XXIX from M31 is quite well-determined to be 207$^{+20}_{-2}$\,kpc, as And XXIX is close to the point of closest approach to M31 along this line of sight (the `tangent point').    

To estimate the luminosity of And XXIX, we computed the luminosity function of RGB stars for And XXIX within 1.5 half-light radii of the center of And XXIX and compared this with the luminosity function of RGB stars in Draco within 1.5 half-light radii, incorporating the distance uncertainties of Draco and And XXIX and shot noise in the star samples by using bootstrapping and Monte Carlo techniques.  The RGB luminosity function appears to be best fit by scaling Draco by a factor of 0.63$\pm0.14$, and assuming a luminosity of Draco to be $M_V = -8.8\pm0.3$ following \citet{mateo98}, we infer a luminosity for And XXIX of $M_V \sim -8.3 \pm 0.4$.  

%% \subsection{Structure}

\begin{figure}
\begin{center}
%% ps2eps --rotate=+ --ignoreBB -f AndXXIX_logr.ps 
\plotone{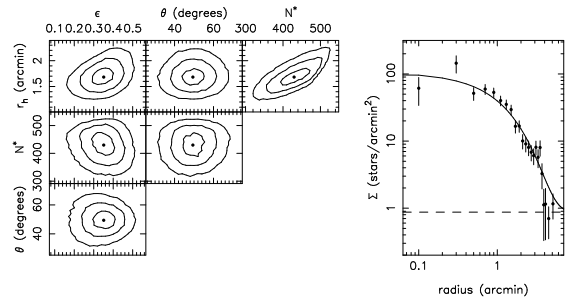}
\caption{{\it Left:} confidence areas for the measurement of half-light radius,
ellipticity, position angle, and number of detected stars. The contours correspond, when projected on the axes, to 1-, 2-, and 3-$\sigma$ uncertainties (to allow reading of the marginalized 1-$\sigma$ value straight from the plot for each parameter).  The filled circles correspond to the peak of the maximum likelihood function.  {\it Right:} the radial profile of And XXIX for which stars have been binned according to the favored structural parameters (points, where the error bars correspond to Poisson uncertainties in the counts of objects consistent with the Gemini GMOS-N point spread function), and compared with the best-fit exponential surface brightness profile (black line).  The measured background level is shown by the dashed, horizontal line. \label{profile_structure}}
\end{center}
\end{figure}

We computed the radial profile of And XXIX, along with the position, half-light
radius, eccentricity, and position angle using the maximum likelihood technique
described by \citet{martin08}, accounting for holes left by the halo of bright stars and galaxies.  We used stars from the GMOS-N imaging with $i<24$ and $0.1<r-i<0.5$ to determine these fit parameters.  
This method assumes an exponential profile for
the dwarf galaxy and a constant background level. Figure~\ref{profile_structure}
shows on the left maximum likelihood contours of the half-light radius,
ellipticity ($\epsilon$), position angle ($\theta$), and number of detected
stars in the overdensity ($N_\star$), while the right side shows the radial
profile fit. The structural parameters have one-dimensional 1-~, 2-~, and
3-$\sigma$ confidence areas overlaid.  
The ellipticity $e = 0.35\pm0.06$ and half-light semi-major axis of 
$r_h = 360 \pm 60$ pc (1.7$^{\prime} \pm 0.2^{\prime}$) are typical of
other Local Group dwarf galaxies at And XXIX's absolute magnitude (see Fig.\ 6 of \citealp{richardson11}, with a range between 200 and 500\,pc half-light radius for both M31 and Milky Way satellites with $-8>M_V>-9$, and Fig.\ 6 of \citealp{martin08}).

\section{Closing Remarks}

In this Letter, we reported the discovery of a new dwarf galaxy, Andromeda XXIX, using data from the recently released SDSS DR8.  The detection in the SDSS data was marginal, and we confirmed the detection using deeper $r$ and $i$-band imaging data from the GMOS instrument on Gemini North.  The GMOS data show a clear RGB and no sign of other bluer luminous stars; accordingly we argue that And XXIX is a dwarf spheroidal galaxy, separated on the sky by a little more than 15 degrees from M31.  The distance modulus inferred from the TRGB is $24.32 \pm 0.22$, corresponding to a heliocentric distance of $730\pm75$\,kpc and a three-dimensional separation from M31 of $207^{+20}_{-2}$\,kpc (close to M31's virial radius).  This separation is close to the distance at which dwarf spheroidal satellites start to become less common, giving way to dwarf irregular satellites (see, e.g., Fig.\ 3 of \citealp{grebel03}).  Its absolute magnitude, as determined by comparison to the RGB luminosity function of the Draco dwarf spheroidal, is $M_V = -8.3\pm 0.4$.  And XXIX's stellar populations appear very similar to Draco's; consequently, we estimate a metallicity for And XXIX of [Fe/H]$\sim -1.8$.  The half-light radius of And XXIX is $360\pm60$\,pc and its ellipticity is $0.35\pm0.06$, typical of dwarf satellites of the Milky Way and M31 at this absolute magnitude range.

\acknowledgments

We thank the referee for their feedback.  This work was partially supported by NSF grant AST 1008342. N.F.M.\ acknowledges funding by Sonderforschungsbereich SFB 881 "The Milky Way System" (subproject A3) of the German Research Foundation (DFG).

Funding for SDSS-III has been provided by the Alfred P. Sloan Foundation, the
Participating Institutions, the National Science Foundation, and the U.S.
Department of Energy.  The SDSS-III Web site is http://www.sdss3.org/. 

SDSS-III is managed by the Astrophysical Research Consortium for the
Participating Institutions of the SDSS-III Collaboration including the
University of Arizona, the Brazilian Participation Group, Brookhaven National
Laboratory, University of Cambridge, University of Florida, the French
Participation Group, the German Participation Group, the Instituto de Astroﬁsica
de Canarias, the Michigan State/Notre Dame/JINA Participation Group, Johns
Hopkins University, Lawrence Berkeley National Laboratory, Max Planck Institute
for Astrophysics, New Mexico State University, New York University, the Ohio
State University, University of Portsmouth, Princeton University, University of
Tokyo, the University of Utah, Vanderbilt University, University of Virginia,
University of Washington, and Yale University. 

Based on observations obtained at the Gemini Observatory, which is operated by the 
Association of Universities for Research in Astronomy, Inc., under a cooperative agreement 
with the NSF on behalf of the Gemini partnership: the National Science Foundation (United 
States), the Science and Technology Facilities Council (United Kingdom), the 
National Research Council (Canada), CONICYT (Chile), the Australian Research Council (Australia), 
Minist\'{e}rio da Ci\^{e}ncia e Tecnologia (Brazil) 
and Ministerio de Ciencia, Tecnolog\'{i}a e Innovaci\'{o}n Productiva (Argentina).

This research has made use of the NASA/IPAC Extragalactic Database (NED) which is operated by the Jet Propulsion Laboratory, California Institute of Technology, under contract with the National Aeronautics and Space Administration (NASA), and NASA's Astrophysics Data System Bibliographic Services.

\clearpage

\begin{thebibliography}{}
\bibitem[Aihara et al.(2011)]{DR8} Aihara, H., et al.\ 2011, \apjs, 193, 29 
\bibitem[Apparicio et al.(2001)]{apparicio01} Apparicio, A., Carrera, R., Martinez-Delgado, D. 2001, AJ 122, 2524
\bibitem[Bellazzini et al.(2001)]{bellazini01} Bellazzini, M., 
Ferraro, F.~R., \& Pancino, E.\ 2001, \apj, 556, 635 
\bibitem[Bramich et al.(2008)]{bramich08} Bramich, D.\ M., et al.\ 2008, \mnras, 386, 887
\bibitem[de Jong et al.(2008)]{dejong08} de Jong, J.\ T.\ A., et al.\ 2008, \apj, 680, 1112
\bibitem[Belokurov et al.(2007)]{belokurov07} Belokurov, V., et al.\ 2007, \apj, 654, 897
\bibitem[Dotter et al.(2008)]{dotter08} Dotter, A., Chaboyer, 
B., Jevremovi{\'c}, D., Kostov, V., Baron, E., 
\& Ferguson, J.~W.\ 2008, \apjs, 178, 89 
\bibitem[Font et al.(2011)]{font11} Font, A.\ S. 2011, \mnras, in press (arXiv:1103.0024)
\bibitem[Grebel et al.(2003)]{grebel03} Grebel, E.~K., Gallagher, J.~S., III, \& Harbeck, D.\ 2003, \aj, 125, 1926 
\bibitem[Hook et al.(2004)]{gmosn} Hook, I.\ M., J{\o}rgensen, I., Allington-Smith, J.\ R., Davies, R.\ L., Metcalfe, N., Murowinski, R.\ G. and Crampton, D.\ 2004, \pasp, 116, 425
\bibitem[Ibata et al.(2007)]{ibata07} Ibata, R., Martin, N.~F., Irwin, M., Chapman, S., Ferguson, A.~M.~N., Lewis, G.~F., \& McConnachie, A.~W.\ 2007, \apj, 671, 1591 
%\bibitem[Irwin et al.(2007)]{irwin07} Irwin, M.~J., et al.\ 2007, \apjl, 656, L13 
\bibitem[Irwin et al.(2008)]{irwin08} Irwin, M.~J., Ferguson, A.~M.~N., Huxor, A.~P., Tanvir, N.~R., Ibata, R.~A., \& Lewis, G.~F.\ 2008, \apjl, 676, L17 
%\bibitem[Jordi et al.(2006)]{jordi06} Jordi, K., Grebel, E.~K., \& Ammon, K.\ 2006, \aap, 460, 339 
\bibitem[Kinemuchi et al.(2008)]{kinemuchi08} Kinemuchi, K., Harris, H.\ C., Smith, H.\ A., Silbermann, N.\ A., Snyder, L.\ A., LaCluyze, A.\ P. \&  Clark, C.\ L. 2008, \aj, 136, 1921
\bibitem[Klypin et al.(1999)]{klypin99} Klypin, A., Kravtsov, A.\ V., Valenzuela, O., \& Prada, F. 1999, \apj, 522, 82
\bibitem[Macci{\`o} et al.(2010)]{maccio10} {Macci{\`o}}, A.~V., {Kang}, X., {Fontanot}, F., {Somerville}, R.~S., {Koposov}, S. \& {Monaco}, P. 2010, \mnras, 402, 1995
\bibitem[Makarov et al.(2006)]{makarov06} Makarov, D., Makarova, L., Rizzi, L., Tully, R.~B., Dolphin, A.~E., Sakai, S., \& Shaya, E.~J.\ 2006, \aj, 132, 2729 
\bibitem[Martin et al.(2006)]{martin06} Martin, N.~F., Ibata, R.~A., Irwin, M.~J., Chapman, S., Lewis, G.~F., Ferguson, A.~M.~N., Tanvir, N., \& McConnachie, A.~W.\ 2006, \mnras, 371, 1983 
\bibitem[Martin et al.(2008)]{martin08} Martin, N.~F., de Jong, 
J.~T.~A., \& Rix, H.-W.\ 2008, \apj, 684, 1075 
\bibitem[Martin et al.(2009)]{martin09} Martin, N.~F., et al.\ 2009, \apj, 705, 758 
\bibitem[Mateo(1998)]{mateo98} Mateo, M. 1998, \araa, 36, 435
% \bibitem[Mayer et al.(2006)]{mayer06} Mayer, L., Mastropietro, C., Wadsley, J., Stadel, J., \& Moore, B.\ 2006, \mnras, 369, 1021 
\bibitem[McConnachie et al.(2005)]{mcconnachie05} McConnachie, A.~W., Irwin, M.~J., Ferguson, A.~M.~N., Ibata, R.~A., Lewis, G.~F., \& Tanvir, N.\ 2005, \mnras, 356, 979 
\bibitem[McConnachie et al.(2009)]{mcconnachie09} McConnachie, A.~W.,  et al.\ 2009, \nat, 461, 66 
\bibitem[Moore et al.(1999)]{moore99} Moore, B., Governato, F., Lake, G., Quinn, T., Stadel, J., \& Tozzi, P. 1999, \apj, 524, L19
\bibitem[Richardson et al.(2011)]{richardson11} Richardson, J.\ C., et al. 2011, \apj, 732, 76
\bibitem[Schlegel et al.(1998)]{SFD98} Schlegel, D.~J., 
Finkbeiner, D.~P., \& Davis, M.\ 1998, \apj, 500, 525 
\bibitem[S\'egall et al.(2007)]{segall06} S\'egall, M., Ibata, R.\, A., Irwin, M.\ J., Martin, N.\ F., \& Chapman, S. 2007, \mnras, 375, 831
\bibitem[Slater et al.(2011)]{slater11} Slater, C.\ T., Bell, E.\ F. \& Martin, N.\ F. 2011, \apj \ Letters, in press
\bibitem[Stetson(1987)]{stetson87} Stetson, P.\ B. 1987, \pasp, 99, 191
\bibitem[Tammann, Sandage \& Reindl(2008)]{tammann08} Tammann, G.\ A., Sandage, A., \& Reindl, B. 2008, \apj, 679, 52
%\bibitem[Thuan \& Martin(1979)]{thuan79} Thuan, T.~X., \& Martin, G.~E.\ 1979, \apjl, 232, L11
\bibitem[Vilardell et al.(2010)]{vil10} Vilardell, F., Ribas, I., Jordi, C., Fitzpatrick, E.\ L. \& Guinan, E.\ F. 2010, \aap, 509, 70
%\bibitem[Weisz et al.(2011)]{weisz11} Weisz, D.~R., et al.\ 2011, arXiv:1101.1093 
\bibitem[Willman et al.(2005)]{willman1} Willman, B., et al.\ 2005, \aj, 129, 2692
\bibitem[York et al.(2000)]{SDSS} York, D.\ G. et al.\ 2000, \aj, 120, 1579
\bibitem[Zucker et al.(2004)]{zucker04} Zucker, D.~B., et al.\ 2004, \apjl, 612, L121 
\bibitem[Zucker et al.(2007)]{zucker07} Zucker, D.~B., et al.\ 
2007, \apjl, 659, L21 

\end{thebibliography}
\end{document}